\documentclass[runningheads]{llncs}
\usepackage{comment}
\usepackage{float}
\usepackage{graphicx}
\usepackage{enumitem}
\usepackage{subcaption}
\begin{document}
\title{RateRL: A Framework for Developing RL-Based Rate Adaptation Algorithms in ns-3}
%
%
\author{Ruben Queiros\inst{1,2}\orcidID{0000-0001-7804-0246} \and
Luís Ferreira\inst{1,2}\orcidID{ 0009-0002-9107-0639} \and
Helder Fontes\inst{2}\orcidID{0000-0002-7672-8335}
\and
Rui Campos\inst{1,2}\orcidID{0000-0001-9419-6670}}

\authorrunning{R. Queiros et al.}
%
\institute{Faculdade de Engenharia da Universidade do Porto, Porto, Portugal \and
INESC TEC, Porto, Portugal 
\email{\{ruben.m.queiros, luis.m.martins, helder.m.fontes, rui.l.campos\}@inesctec.pt}}
\maketitle              
\begin{abstract}

The increasing complexity of recent Wi-Fi amendments is making the use of traditional algorithms and heuristics unfeasible to address the Rate Adaptation (RA) problem. This is due to the large combination of configuration parameters along with the high variability of the wireless channel. Recently, several works have proposed the usage of Reinforcement Learning (RL) techniques to address the problem. However, the proposed solutions lack sufficient technical explanation. Also, the lack of standard frameworks enabling the reproducibility of results and the limited availability of source code, makes the fair comparison with state of the art approaches a challenge. This paper proposes a framework, named RateRL, that integrates state of the art libraries with the well-known Network Simulator 3 (ns-3) to enable the implementation and evaluation of RL-based RA algorithms. To the best of our knowledge, RateRL is the first tool available to assist researchers during the implementation, validation and evaluation phases of RL-based RA algorithms and enable the fair comparison between competing algorithms.

\end{abstract}

\section{Introduction}

The new configuration parameters available in the most recent Wi-Fi amendments allied to the high variability and asymmetry of the radio channel, make Rate Adaptation (RA) and the optimal configuration of these parameters challenging. Recent RA algorithms in Wi-Fi are often developed considering simplified simulations that are not always well described. This poses a challenge for the implementation and comparison of different RA algorithms. For example, the Network Simulator 3 (ns-3) implements a significant amount of existing state-of-the-art RA algorithms. However, only a few of these algorithms can be used for recent Wi-Fi versions subsequent to IEEE~802.11n. Additionally, many recent RA algorithms are based on Machine Learning (ML), which can make them difficult to implement and understand. In some cases, the authors of these algorithms do not even provide the source code or the training dataset, which poses an obstacle to accurately reproduce the obtained results. When developing novel ML-based solutions the authors should consider the following practices: 1) clearly describe the problem that the proposal is trying to solve and justify any underlying assumptions; 2) systematically devise procedures that ensure the reproducibility of results; 3) clearly state and justify the metrics used for the evaluation; 4) expose the dataset used in the training process of the algorithm; and 5) have one or more baseline models to compare with.

The use of Reinforcement Learning (RL) and other ML techniques within the wireless networks research area has emerged as a way to further improve the network Quality of Service (QoS). Leveraging these techniques, the research community has been optimising the network performance, reducing latency, and ensuring efficient resource allocation~\cite{wifi-meets-ml,ml-improv}. A high percentage of the current telecommunications infrastructures have begun to invest and test ML algorithms for supporting the network operation and business decisions~\cite{9357376}. However, the techniques employed are not standardized, resulting in a lack of foundational principles that would enable the transfer of knowledge from previous initiatives to more recent ones. When compared with other fields that pioneered the use of ML, such as computer vision or natural language processing, wireless networks research started using ML at a later stage. Thus, it is common within the existing literature to identify ML-based solutions that lack the use of good practices stated above. A predominant challenge relates to the scarcity of technical details provided in published works, which hampers the reproducibility of results. Furthermore, the limited availability to the source code and the dataset used, precludes the validation and extension of prior findings. These issues collectively emphasize the need for a more cohesive and systematic approach to integrate RL into wireless networks.

The main contribution of this paper is RateRL, a framework to support the development of RL-based RA algorithms. We illustrate the use of RateRL through a practical use case, employing the Data Driven Algorithm for Rate Adaptation (DARA) proposed in~\cite{dara}. Our framework integrates well-known RL libraries, such as TensorFlow Agents and OpenAI Gym~\cite{openai-gym} with ns-3~\cite{ns3}, using ns3-gym~\cite{ns3gym} to interface all the components. RateRL enables the task automation to identify the hyperparameter configuration that maximizes the expected cumulative reward of the RL algorithm, while offering real-time feedback of the training process. The code and ns-3 scripts are publicly available, facilitating the efforts of future research works to build upon RateRL.

The rest of the paper is organised as follows. Section II presents the background. Section III addresses the related work. Section IV explains the RateRL framework. Section V illustrates the use of RateRL considering DARA as the use case. Finally, Section V provides some concluding remarks and points out the future work.

\section{Background}

This section refers to some background concepts that are necessary to understand RateRL. It starts with an overview of RL, namely the Q-learning algorithm and existing frameworks to implement these techniques. Then, a brief description of the network simulator ns-3 is provided.

\subsection{Reinforcement Learning}

\begin{figure}
    \centering
    \includegraphics[width=.7\linewidth]{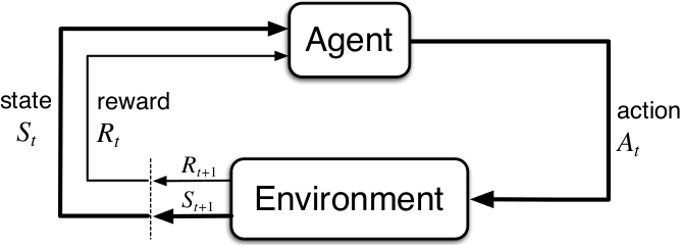}
    \caption{Reinforcement Learning loop diagram.}
    \label{fig:RLdiag}
\end{figure}

The RL model, represented in Figure~\ref{fig:RLdiag}, is composed of the agent and the environment elements. The communication between them is based on state, action and reward signals. The agent learns the decisions (actions) based on the observations (states) received from the environment and its decision is evaluated taking into account the reward. The objective of RL is to learn a policy of actions to take for a given environment state that maximises the overall cumulative reward for every (state, action) pair. There are multiple learning algorithms available in the literature. The recent works that address the RA problem~\cite{dara,rlcsma,rlccnc,edra} use the classic Q-learning algorithm and show better results when compared to traditional algorithms such as Minstrel-HT~\cite{minstrelht}, the default RA algorithm for Linux systems.

Q-learning~\cite{qlearning} is a model-free algorithm, which means that it does not use any prior knowledge of the environment and learns through trial and error. Q-learning only works with a discrete action space. The objective of Q-learning is to learn the optimal policy that maximises the expected cumulative reward. Q-learning learns and updates its Q-function values via trial and error using Eq.~\ref{eq:qlearning}, where $Q(s,a)$ is the expected cumulative reward when the agent selects action $a$ in state $s$ considering that future actions are selected according to the learnt policy. $r(s,a)$ is the reward for taking action $a$ in state $s$, and $\max_{a \in A}Q(s_{new},a), \forall a \in A$ is the maximum possible reward of the new state $s_{new}$, which is the result of the current action; $s_{new}$ is the new state and $a_{all}$ represents every $a \in A$. The learning rate $\alpha$ determines the rate at which new values update the total Q-value. Finally, the discount factor $\gamma \in [0,1]$ determines the importance of future rewards in the calculation of the expected cumulative reward.

\begin{equation}
    Q(s,a)\leftarrow (1-\alpha)Q(s,a)+\alpha[r(s,a)+\gamma\max_{\forall a \in A}Q(s_{new},a)]
    \label{eq:qlearning}
\end{equation}

\subsection{Reinforcement Learning Frameworks}
There are multiple and extensively validated frameworks available to implement Q-learning or other RL-based learning algorithm. TensorFlow and PyTorch are two prominent deep learning frameworks widely adopted for building and training ML models, including RL agents. TensorFlow, developed by Google, offers a comprehensive ecosystem of tools and libraries that facilitate the creation of neural networks and the optimisation of their parameters. TensorFlow Agents (TF-Agents) is an extension of TensorFlow specifically designed to enable the construction of RL agents. TF-Agents~\cite{TFAgents} provides reusable components for defining agent behaviour, defining the environment interaction loop, and implementing various state-of-the-art RL algorithms. Similarly, PyTorch, developed by Facebook's AI Research lab, provides a dynamic computational graph that simplifies model development and experimentation. Gym is an open-source toolkit developed by OpenAI, designed to facilitate the development and testing of RL algorithms. It provides a collection of environments (simulated scenarios) that allow researchers and developers to experiment with and benchmark various RL techniques as well as to develop their own custom environment making it a valuable resource that can interface with the popular network simulator ns-3.

\subsection{The Network Simulator 3}
ns-3~\cite{ns3} is a well-known open-source network simulation tool and one of the most used wireless network simulators. ns-3 was driven by a desire to model networks in a way that best suits network research and learning. It aims to provide highly accurate and scalable network simulation capabilities for studying various aspects of networking protocols, communications methods, and network behaviour. Hence, it was not developed with ML or artificial intelligence in mind. There is currently no official framework for integrating with prevalent ML tools. Initial community efforts~\cite{ns3gym,ns3ai} were made to bridge the gap between the network simulator and popular ML frameworks such as TensorFlow, PyTorch and Gym. The resulting tools, named ns3-ai and ns3-gym, are detailed in Section~\ref{sec:rw}.

\section{Related Work}\label{sec:rw}

The related work is presented in this section. First, we overview existing reinforcement learning frameworks for networking and then we review the state of the art in RL-based solutions for the RA problem.

\subsection{Reinforcement Learning Frameworks for Networking}

To the best of our knowledge, few frameworks that integrate Reinforcement Learning frameworks with network simulators have been proposed in the state of the art. In~\cite{ns3gym} the authors develop a sockets-based interface between ns-3 and OpenAI Gym to encourage the usage of RL in networking research. To overcome the lack of flexibility imposed by ns3-gym, the authors in~\cite{ns3ai} proposed ns3-ai, a Python module that allows the interconnection between any artificial intelligence framework with ns-3, using a higher efficiency mechanism based on shared-memory. However, since this is a high flexibility and efficiency data interaction framework researchers using ns3-ai have to adapt to implement their own development environment to address their specific problem, resulting in slower algorithm development or simplistic simulation environments, which reduces the solution realism. On the other hand, the authors in~\cite{prisma} developed the PRISMA framework, which is tailored to the distributed packet routing problem, on top of ns3-gym, extending the problem to a Multi-Agent Deep Reinforcement Learning approach. Therefore, researchers would need to modify PRISMA's design if the objective was to address the RA problem.

\subsection{Reinforcement Learning for Rate Adaptation}

Different solutions have been proposed to solve the RA problem, in particular for Wi-Fi networks. Some use classical heuristic approaches~\cite{minstrelht,strale} while recent works have been using RL-based techniques~\cite{smartla,edra,rlcsma,rlccnc,dara}. In~\cite{edra}, the authors developed a Q-learning based link adaptation solution that addresses RA together with other configuration parameters such as the channel bandwidth and number of spatial streams, outperforming state of the art solutions in terms of throughput. Despite implementing it in a network interface card and evaluating their solution in an experimental setting, the authors do not mention any standard framework that was used to implement their solution. Also, the source code is not publicly available. In~\cite{rlcsma,rlccnc} the authors use ns-3 to implement their RL-based RA solutions with the help of ns3-gym and ns3-ai frameworks. However, both works do not provide any information with regards to the training process or hyperparameter configuration. Moreover, despite simplistic simulation scenarios for evaluation, they are not sufficiently well described, posing an obstacle to accurately reproduce the obtained results.

\subsection{Summary}

Despite the good results of recent works that boost the overall network performance, the difficulty to replicate the results of the proposed solutions is common among existing works. Typically, they lack implementation details -- i.e. the code is not open source -- and training process description. Moreover, within the identified works, the results of the proposed solutions are not compared with other RL-based RA algorithms.

\section{RateRL Framework}

In this section, we present the RateRL framework, including its architecture and components.

The RateRL framework was designed considering design principles similar to the PRISMA framework~\cite{prisma}, such as the achievement of realistic wireless networks simulation environments due to the usage of ns-3 and the development with a modular approach, which makes fast prototyping of RL-based RA algorithms possible.

\begin{figure}
    \centering
    \includegraphics[width=\linewidth]{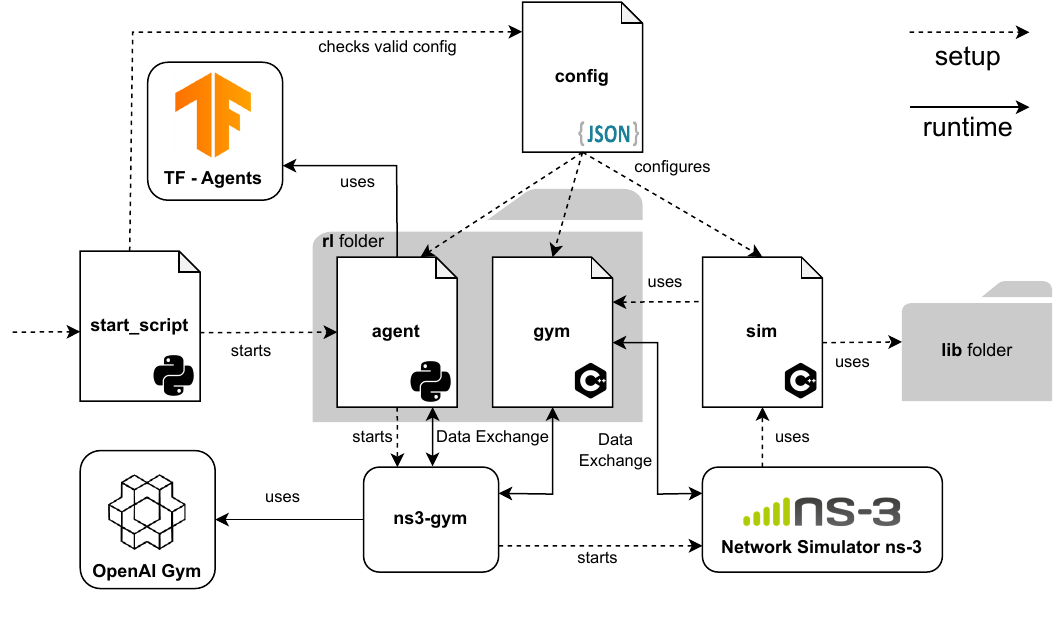}
    \caption{RateRL architecture diagram.}
    \label{fig:RateRL-arch}
\end{figure}

In Figure~\ref{fig:RateRL-arch} we present the RateRL architecture and how its components interact with each other. The programming languages used to code RateRL were Python and C++. The Python files hold every essential setup procedures and parsing mechanisms required for the proper execution of the system, the core ML components and most of the data collection processes. The files written in C++, are mainly related to the configuration and utilisation of the ns-3 simulator together with some files that interact with the ns3-gym interface.    

We now detail the RateRL components and instruct the reader on how to use RateRL. The starting point of the RateRL framework is the \texttt{start\_script} file. A detailed description of the role of each component is presented below:

\begin{itemize}
    \item{\texttt{start\_script}} checks if the configuration file is valid. It also sets up the results folder where the logs of the simulation or training are stored. Finally, it starts the agent component. 
    \item{\texttt{config}} is a JSON file that holds the configuration parameters relevant for each of the other files within the framework, such as the \texttt{agent}, \texttt{gym} and \texttt{sim}.
    \item{\texttt{agent}} interacts with the TF-Agents framework using the environment data that is retrieved from the ns-3 simulation. It starts the ns-3 environment using the ns3-gym interface.
    \item{\texttt{sim}} defines the ns-3 simulation script. It uses the \texttt{lib} folder, where multiple utility functions are defined to ease aspects such as the simulation configuration and data collection. 
    \item{\texttt{gym}} configures the ns-3 environment defining the observation and action data shapes. It executes the action that comes from the \texttt{agent}, and establishes the data collection methods that are used to collect the observation from the ns-3 simulation as well as the metrics to calculate the reward.  
\end{itemize}

The installation process is documented in the public repository, inside the \texttt{install} folder, and it assumes the user does not have any of the required dependencies installed. The user can run the agent in two different modes: 

\begin{itemize}
    \item \textbf{Training Mode} starts by filling the replay buffer with the trajectories collected from the simulation. A trajectory consists of a time step (i.e., the initial observation of the environment), the action step (i.e., the action that was taken considering the previous time step), and the next time step (i.e., the new observation and the reward that was obtained using the previous action step). This replay buffer is filled while the simulation is running until the end of the episode, which in this case is the end of the simulation. The agent is trained during this process, grabbing randomly from the replay buffer an amount of trajectories defined with the hyperparameter \textit{batch size}, updating the weights accordingly and increasing the train step counter. The user can then adapt the amount of episodes the simulation runs, how frequent the training happens and the way epsilon greedy adjusts over the training process. When the training is over, or if the user wants to pause it, it is possible to save the progress with a checkpoint so that the current state of the policy can be recovered later.
    \item \textbf{Evaluation Mode} loads the trained Policy and assumes a fixed epsilon greedy factor of 0 to avoid exploratory attempts. However, this mode is not prepared for simulation scenarios that dynamically change requiring an online learning approach. This will be the subject of future work.
\end{itemize}

Regardless of the mode used, RateRL saves simulation logs with the throughput of every existing communication link as well as the nodes positions, with a configurable periodicity.

\section{Using the RateRL Framework}
In this section we use the implementation, training and evaluation of the DARA algorithm~\cite{dara} as a use case to illustrate the utilisation of the RateRL framework. 

\subsection{DARA Overview}

DARA~\cite{dara} is a RL-based RA algorithm developed for the IEEE 802.11n amendment. It considers scenarios with Single Input Single Output and fixed channel bandwidth of 20 MHz, using long Guard Interval. The valid actions are the first 8 Modulation and Coding Schemes (MCS). The state is the average SNR value considering the Acknowledgement frames that originate from the receiver node. Finally, the reward is a function of the Frame Success Ratio (FSR) and the chosen MCS, to value the highest possible MCS without compromising the FSR.

\subsection{Simulation Settings}

We configured the preliminary validation scenario defined in~\cite{dara}, . In this scenario, we have a stationary node and a moving node. In the beginning of the simulation, the nodes start close to each other, and their distance increases throughout the simulation period. In this way we stimulate the algorithm with a wide range of SNR values. The algorithm is then compared in terms of throughput with other RA algorithms implemented in ns-3 such as Minstrel-HT (MIN) and the Ideal (ID) algorithm. All the other main simulation configuration parameters are presented in Table~\ref{tab:ns3config}.

\begin{table}
\centering
\caption{Simulation Configuration Parameters.}
\begin{tabular}{ll}
\hline
\textbf{Configuration Parameter} & \textbf{Value}                                      \\ \hline
Wi-Fi Standard          & IEEE 802.11n                               \\
Propagation Delay Model & Constant Speed                             \\
Propagation Loss Model  & Friis                                      \\
Frequency               & 5180 MHz                                   \\
Channel Bandwidth       & 20 MHz                                     \\
Transmission Power      & 20 dBm                                     \\
Wi-Fi MAC               & Ad-hoc                                     \\
Traffic                 & UDP, generated above link capacity         \\
Packet Size             & 1400 Bytes of UDP Payload                  \\
\hline
\end{tabular}
\label{tab:ns3config}
\end{table}

\subsection{Training and Hyperparameter Tuning}

DARA was trained and evaluated on a ASUS ROG G14 Laptop with a Ryzen 9 5900HS (8 cores up to 4.6 GHz), 32 GB RAM and a NVIDIA RTX 3060 GPU. In this illustrative example the hyperparameter configuration chosen is defined in Table~\ref{dqnlearn}. The simulations were 60 seconds long, for a total of 15 episodes. 

\begin{table}
\centering
\caption{DQN Learning Algorithm main Parameters.}
\begin{tabular}{p{0.4\linewidth}p{0.5\linewidth}}
\hline
\textbf{Parameter} & \textbf{Value}                               \\
\hline
Observation Space       & One-dimensional scaled float (0.0-1.0)  \\
Action Space            & One-dimensional integer (0-7)           \\
Optimiser               & Adam                                    \\
Loss Function           & Mean Square Error                       \\
Epsilon Greedy          & Fixed at 0.1                            \\
Discount Factor         & Fixed at 0.5                            \\   
Replay buffer           & size of $10^{6}$                        \\
Batch Size              & 64                                      \\
\hline
\end{tabular}
\label{dqnlearn}
\end{table}

The hyperparameter tuning is an essential part of any machine learning model training. To this end, we assess how different values could benefit the final DARA performance. In this work, comparisons between different values of learning rate and the hidden layer architecture of the neural network were carried out. 

These comparisons were not extensive, thus the performance of DARA could be further improved with a more in depth tuning, despite the simple scenario, which shows that good results are achievable with few training episodes. However, the objective of this work is to show that RateRL can be used to compare different hyperparamenter configurations and assess their impact in the performance of the algorithm.

\begin{figure}
    \centering
    \begin{subfigure}{0.45\textwidth}
    \includegraphics[trim={0cm 0cm 0cm 1cm}, clip, width=\textwidth]{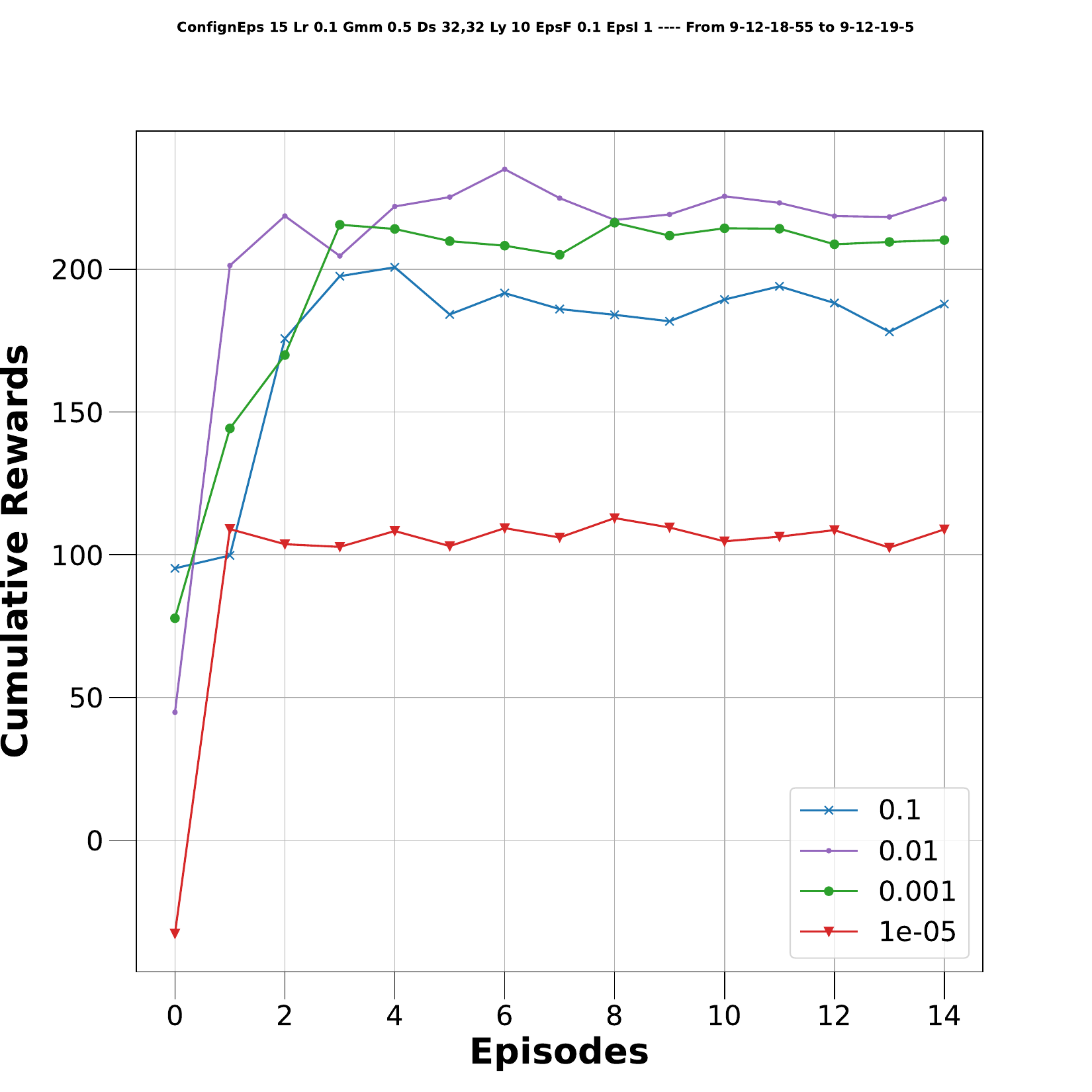}
    \caption{Learning Rate comparison training.}
    \label{fig:lr}
    \end{subfigure}
    \begin{subfigure}{0.45\textwidth}
    \includegraphics[trim={0cm 0cm 0cm 1cm}, clip, width=\textwidth]{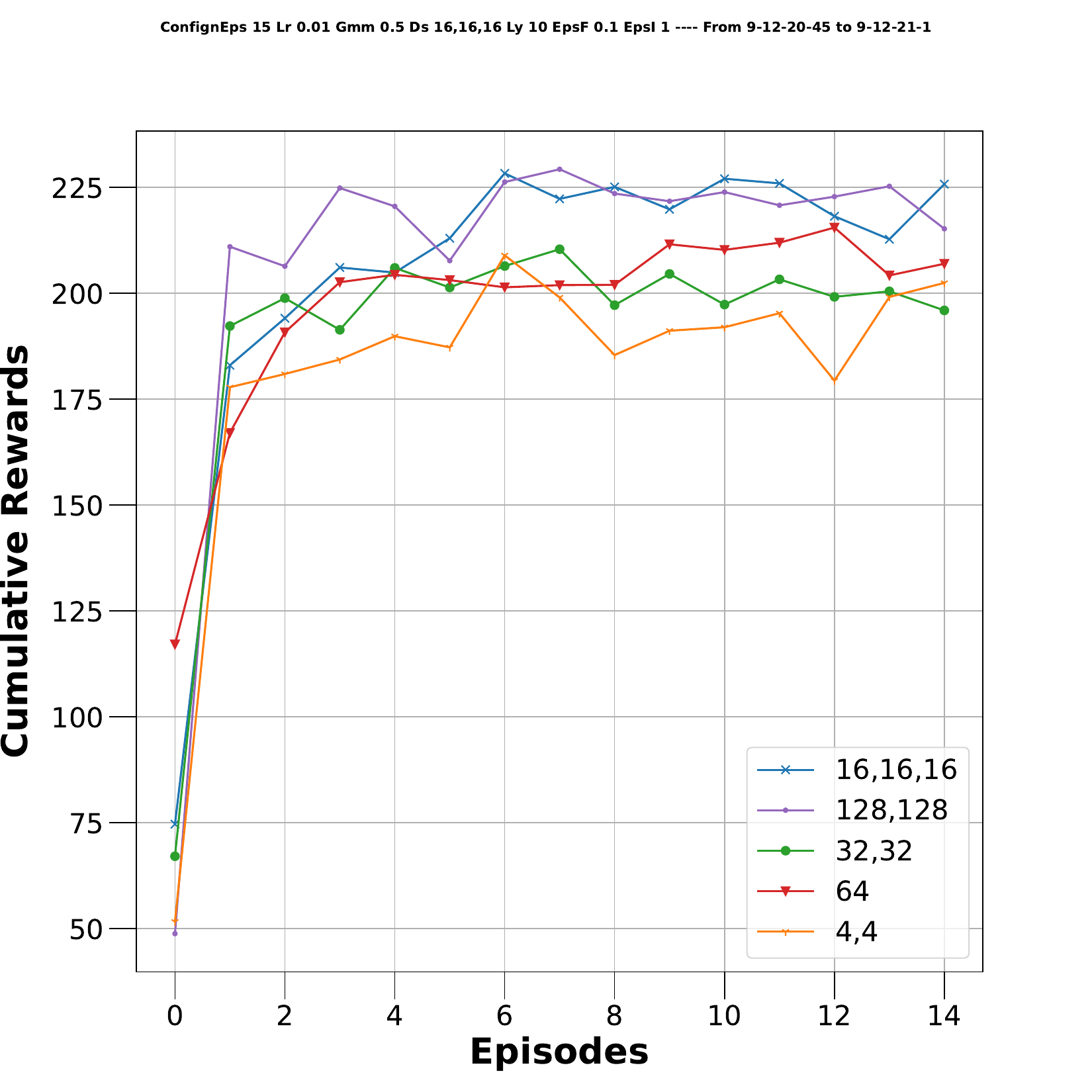}
    \caption{Hidden Layer architecture training.}
    \label{fig:hl}
    \end{subfigure}
    \caption{Hyperparameter tuning trainings.}
    \label{abc}
    \end{figure}

Figure~\ref{fig:lr} shows the cumulative reward over 15 training episodes with 4 different learning rate configurations, using two hidden layers with 32 units each. The results show that a learning rate of 0.01 is consistently better than the other options. After defining the used learning rate value, additional trainings were performed to fine tune the hidden layer architecture.

Figure~\ref{fig:hl} shows the results of this training, with 5 different options being evaluated. Despite the similarities in performance it was decided to choose as the final configuration the one which finished with highest cumulative reward by the end of the 15 episodes training. Therefore, a learning rate of 0.01 and a neural network with 3 hidden layers of 16 units each was defined.

\subsection{Simulation Results}

\begin{figure}
    \centering
    \begin{subfigure}{0.45\textwidth}
    \includegraphics[trim={0cm 0cm 0cm 0cm}, clip, width=\textwidth]{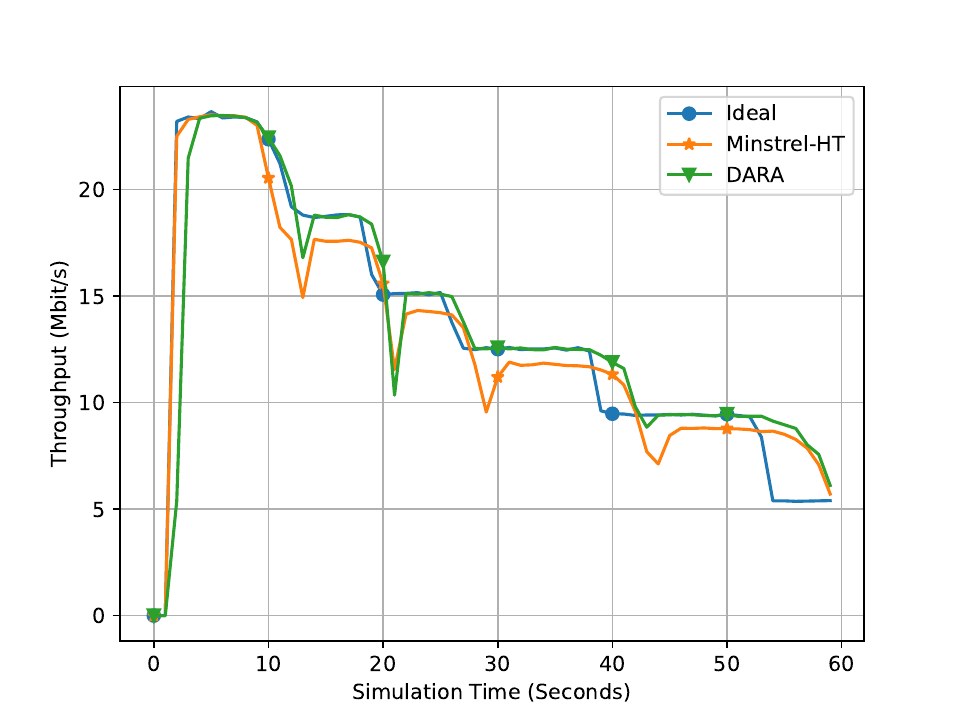}
    \caption{Throughput}
    \label{fig:simthp}
    \end{subfigure}
    \begin{subfigure}{0.45\textwidth}
    \includegraphics[trim={0cm 0cm 0cm 0cm}, clip, width=\textwidth]{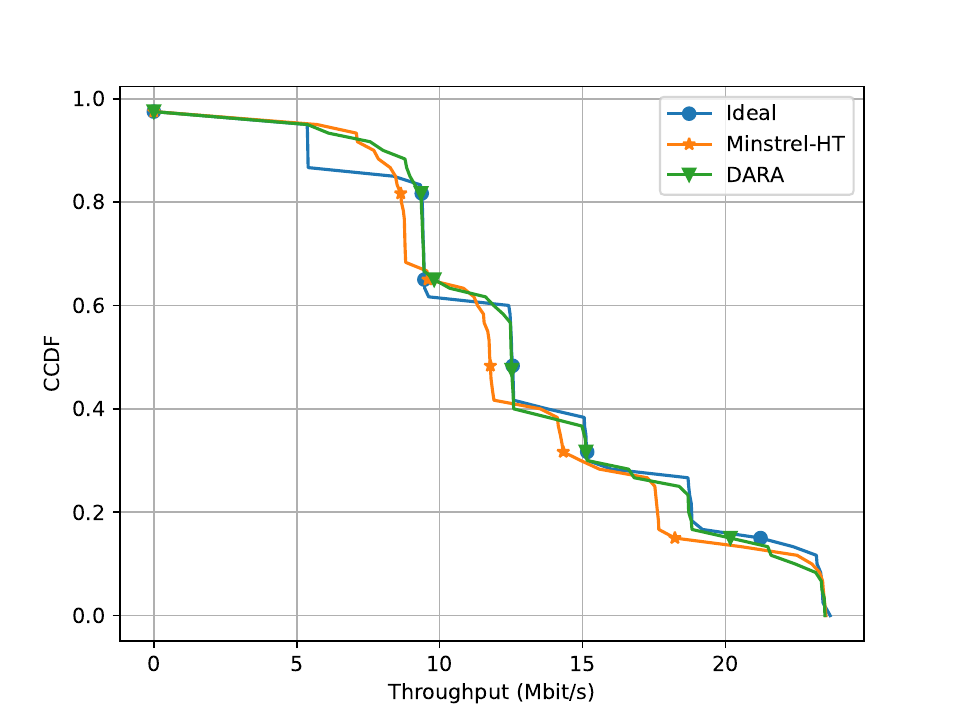}
    \caption{Throughput CCDF}
    \label{fig:simccdf}
    \end{subfigure}
    \caption{Simulation Results using the resulting hyperparamenter configuration.}
    \label{fig:simres}
    \end{figure}

Using the resulting policy from the training that was detailed in the previous sections, we used RateRL to evaluate how the performance of DARA is compared to other popular RA algorithms such as Minstrel and Ideal. Fig~\ref{fig:simthp} shows the throughput throughout the simulation period and  Fig~\ref{fig:simccdf} its complementary cumulative distribution function.

The results show that RateRL can be used to evaluate the performance and comparison of RL-based RA algorithms with other state of the art RA solutions. The average throughput of Ideal was of 13.45 Mbit/s and Minstrel was of 13.07 Mbit/s. DARA achieved an average throughput of 13.52 Mbit/s, an increase of 3.4\% over Minstrel and similar throughput when compared with Ideal. To conclude, we managed to successfully implement train and evaluate DARA using RateRL.

\section{Conclusion}

This paper presented RateRL, the first framework designed for assiting the development, validation and evaluation of RL-based RA algorithms. We demonstrated the use of RateRL in the whole development cycle of an RL-based RA algorithm, using the state of the art DARA algorithm as a use case. Our objective with RateRL is to provide a framework for developing future RL-based RA solutions and enable their direct comparison with state of the art or related solutions. The RateRL framework is open source and it is publicly available on Gitlab~\footnote[1]{https://gitlab.inesctec.pt/pub/ctm-win/raterl}.

As future work, we plan to migrate to ns3-ai to support other popular ML frameworks. Also, we aim to extend RateRL to use other popular RL algorithms such as Deep Deterministic Policy Gradient and Proximal Policy optimisation.

\section*{Acknowledgements}
This work is financed by National Funds through the Portuguese funding agency, FCT - Fundação para a Ciência e Tecnologia, under the PhD grant 2022.10093.BD.

%
%
%
\bibliographystyle{splncs04}
\bibliography{refs.bib}
\end{document}